\documentclass{PoS}
\usepackage{epsfig}
\usepackage{graphicx}
\usepackage[square,sort&compress,comma,numbers]{natbib}
\usepackage{pifont}
\usepackage{latexsym,times}
\usepackage{amssymb,amsmath}
\usepackage{amsxtra}
\usepackage{amscd}
\usepackage{amsthm}
\usepackage{amsfonts}
\usepackage{dsfont}
\usepackage{xcolor}
\usepackage{cancel}
\usepackage{framed}
\usepackage{enumitem}
\usepackage{float}
\usepackage{caption}
\newcommand\So{S_{\scriptscriptstyle{{(0)}}}}
\newcommand\To{\EM_{\scriptscriptstyle{{(0)}}}}
\newcommand\rc{r_{\rm{c}}}

\newcommand{\half}{{{\textstyle\frac{1}{2}}}}
\newcommand{\quarter}{{{\textstyle\frac{1}{4}}}}
\newcommand{\be}{\begin{equation}}
\newcommand{\ee}{\end{equation} }
\newcommand{\beqa}{\begin{eqnarray} }
\newcommand{\eeqa}{\end{eqnarray} }
\newcommand{\ba}{\begin{array}}
\newcommand{\ea}{\end{array}}
\newcommand{\bpm}{\begin{pmatrix}}
\newcommand{\epm}{\end{pmatrix}}

\newcommand{\Spin}{\mathbf{Spin}}

\newcommand{\rmd}{{\rm d}}

\newcommand{\ODD}{\mathbf{O}(D,D)}

\newcommand{\SpinD}{{\Spin(1,D{-1})}}
\newcommand{\oSpinD}{{{\Spin}(D{-1},1)}}

\newcommand{\Spint}{{\Spin(1,9)}}
\newcommand{\oSpint}{{{\Spin}(9,1)}}

\newcommand{\eleven}{{(11)}}
\newcommand{\gammae}{\gamma^{\eleven}}

\newcommand{\brcF}{\bar{\cF}}

\newcommand\EM{T}
\newcommand\EK{K}

\newcommand\mPhi{m_{\scriptscriptstyle{\Phi}}}
\newcommand\mpsi{m_{\scriptscriptstyle{\psi}}}

\newcommand{\DFT}{\rm{DFT}}

\newcommand\Tr{{\rm Tr}}

\newcommand\rd{{\rm d}}

\newcommand\cA{{\cal A}}

\newcommand\cC{{\cal C}}
\newcommand\cD{{\cal D}}

\newcommand\cF{{\cal F}}

\newcommand\cH{{\cal H}}

\newcommand\cJ{{\cal J}}

\newcommand\cL{{\cal L}}

\newcommand\cV{{\cal V}}

\newcommand\hcL{{\hat{\cal L}}}

\newcommand\rhop{{\rho^{\prime}}{}}

\newcommand\psip{\psi^{\prime}}

\newcommand\brrhop{\brrho^{\prime}{}}
\newcommand\brpsip{\brpsi^{\prime}}

\newcommand\dis{\displaystyle}

\def\tx{\tilde{x}}

\def\tpartial{\tilde{\partial}}

\def\bra{\bar{a}}

\def\bri{\bar{\imath}}

\def\breta{\bar{\eta}}
\def\bralpha{\bar{\alpha}}
\def\brbeta{\bar{\beta}}
\def\brgamma{\bar{\gamma}}

\def\brrho{\bar{\rho}}
\def\brpsi{\bar{\psi}}

\def\brl{{\bar{l}}}

\def\brn{{\bar{n}}}
\def\brp{{\bar{p}}}
\def\brq{{\bar{q}}}

\def\brPhi{{{\bar{\Phi}}}}

\def\brC{\bar{C}}

\def\brV{{\bar{V}}}

\def\brX{{\bar{X}}}
\def\brY{{\bar{Y}}}
\def\brP{{\bar{P}}}

\def\brtheta{\bar{\theta}}

\newcommand{\na}{{\nabla}}
\newcommand{\trd}{{\bigtriangledown}}

\newcommand{\red}[1]{{\color{red} #1 \color{black}}}

\newcommand{\blue}[1]{{\color{blue} #1 \color{black}}}

\title{ $\mathbf{O}(D,D)$ completion of  the Einstein Field  Equations}

\ShortTitle{\blue{Einstein Double Field Equations,} $\red{G_{AB}=8\pi G T_{AB}}$ }

\author{\speaker{Jeong-Hyuck Park}~\\~\\
        Department of Physics, Sogang University, 35 Baekbeom-ro, Mapo-gu, Seoul 04107, Korea \\
        E-mail: \email{park@sogang.ac.kr}}

\abstract{
Upon treating the whole closed-string     massless NS-NS sector as stringy graviton fields, {Double Field Theory may evolve into `Stringy Gravity'}. In terms of  an $\mathbf{O}(D,D)$ covariant   differential geometry beyond Riemann, we present  the definitions of the off-shell conserved  stringy Einstein curvature  tensor  and the on-shell conserved  stringy Energy-Momentum tensor. Equating them, all the equations of motion of the  massless   sector are unified into a single expression, $\blue{G_{AB}{=8\pi G} T_{AB},}$ carrying  $\mathbf{O}(D,D)$ vector indices,   
which we dub \blue{{\textit{the Einstein Double Field Equations.}}}

\begin{center}
\vspace{12pt}
\textbf{Proceeding based on a  work~[1] with Stephen Angus and  Kyoungho Cho}\\
~\\
--- {{\textit{{\,Contents\,}}}} ---\\\vspace{3pt}
 \textit{1. Core Idea}\\\vspace{2pt}
 \textit{2. DFT as Stringy Gravity}\\\vspace{2pt}
 \textit{3. Derivation of the Einstein Double Field Equations}\\\vspace{2pt}
 \textit{4. DFT as Modified Gravity}\\
\vspace{30pt}
{\underline{\textbf{To the memory of Cornelius Sochichiu}}}
\end{center}
}

\FullConference{Corfu Summer Institute 2018 "School and Workshops on Elementary Particle Physics and Gravity"\\
		(CORFU2018)\\
		31 August - 28 September, 2018\\
		Corfu, Greece}

\begin{document}

\section{Core Idea}
\noindent String theory  may predict its own gravity rather than General Relativity. In GR, the metric is the only geometric and gravitational field, whereas in string theory the closed-string   massless NS-NS sector comprises a skew-symmetric $B$-field and the string dilaton in addition to the Riemannian metric. $\mathbf{O}(D,D)$ T-duality  rotations   transform them  into each  other~\cite{Buscher:1987sk,Buscher:1987qj}. This hints at a natural augmentation of GR: upon treating the whole closed-string massless NS-NS sector as stringy graviton fields, {Double Field Theory (DFT)~\cite{Siegel:1993xq,Siegel:1993th,Hull:2009mi,Hull:2009zb,Hohm:2010pp} may evolve into Stringy Gravity}.  In terms of   an $\mathbf{O}(D,D)$ covariant stringy  differential geometry beyond Riemann, or the  so-called  semi-covariant formalism~\cite{Jeon:2010rw,Jeon:2011cn},  we  present  the definitions of the off-shell conserved  stringy Einstein curvature tensor~\cite{Park:2015bza} and the on-shell conserved  stringy Energy-Momentum tensor~\cite{Angus:2018mep}. Equating them as prescribed by the action principle of DFT coupled to generic matter, all the equations of motion of the closed string massless NS-NS sector are unified into a single expression,
\be
{G_{AB}=8\pi G T_{AB}\,,}
\label{EDFE}
\ee
which carry  $\mathbf{O}(D,D)$ vector indices. As they correspond to the    $\mathbf{O}(D,D)$ completion of the  (undoubled) Einstein Field  Equations,     we dub them  {\textit{the Einstein Double Field Equations}}~\cite{Angus:2018mep}.\\


\section{{DFT as  Stringy Gravity \,--\, \textit{Essential Constituents}}}

 {$\bullet$ \textbf{Built-in  symmetries \&  Notation:}} 
 \begin{itemize} 
\item[--] $\ODD$ T-duality  
\item[--] DFT diffeomorphisms\, (ordinary diffeomorphisms plus $B$-field gauge symmetry)
\item[--]Twofold local Lorentz symmetries,     $\SpinD\times\oSpinD$
\item[] $\Rightarrow$  Two  locally inertial frames exist  separately  for the left and the right modes.
\end{itemize}
\vspace{7pt}
\begin{center}
\begin{tabular}{c|c|c}
\hline
~~~Index~~~&~~Representation~~&~~Metric (raising/lowering indices)~~\\
\hline
~$A,B,\cdots,M,N,\cdots~$~&$\ODD$   vector&$\cJ_{AB}=\scriptscriptstyle{\bf\left(\ba{cc}0&1\\1&0\ea\right)}$\\
$p,q,\cdots$~&~$\SpinD\,$  vector~~&~$\eta_{pq}=\mbox{diag}(-++\cdots+)$ \\
$\alpha,\beta,\cdots$~&~$\SpinD\,$  spinor~~&
\quad$C_{\alpha\beta},$\quad\quad$(\gamma^{p})^{T}=C\gamma^{p}C^{-1}$~\\
$\brp,\brq,\cdots$~&~$\oSpinD\,$  vector~~&~$\breta_{\brp\brq}=\mbox{diag}(+--\cdots-)$ \\
$\bralpha,\brbeta,\cdots$~&~$\oSpinD\,$  spinor~~&
\quad$\brC_{\bralpha\brbeta},$\quad\quad$(\brgamma^{\brp})^{T}=
\brC\brgamma^{\brp}\brC^{-1}$~\\
\hline
\end{tabular}
\end{center}
~\\
\qquad The   $\ODD$   metric  $\cJ_{AB}$ divides  doubled coordinates into two:    
$x^{A}=(\tx_{\mu},x^{\nu}), \partial_{A}=(\tpartial{}^{\mu},\partial_{\nu})$.\\
\\
{$\bullet$ \textbf{Doubled-yet-gauged spacetime:}}\\
The doubled coordinates are \textit{gauged} through a certain equivalence relation~\cite{Park:2013mpa},
\[
\ba{ll}
x^{A}\,\sim\,x^{A}+\Delta^{A}\,,\qquad&\qquad
\Delta^{A}=\Phi\partial^{A}\Psi\,,
\ea
\] 
where, with $\partial^{A}=\cJ^{AB}\partial_{B}$,   $\,\Delta^{A}$ is derivative-index-valued for arbitrary functions, $\Phi,\Psi$,  appearing in DFT.   Each equivalence class, or gauge orbit in ${{\mathbb{R}}}^{D+D}$,  then corresponds to a single physical point in ${{\mathbb{R}}}^{D}$. This implies, and  also is  implied by, a section condition, 
\[
\partial_{A}\partial^{A}=0\,,
\] 
which can be conveniently solved by switching off the tilde-coordinate dependence,  \textit{i.e.~}$\tpartial^{\mu}\equiv 0$. \\

\noindent   In fact, if we   gauge    the infinitesimal coordinate one-form, $\rmd x^{A}$, explicitly  introducing  a derivative-index-valued auxiliary gauge potential,
\[
\rmd x^{A}\quad\longrightarrow\quad {\rm{D}}x^{A}=\rmd x^{A}-\cA^{A}\,,\qquad\qquad\cA^{A}\partial_{A}=0\,,
\]
it is possible to define an $\ODD$ and DFT-diffeomorphism  covariant `proper length' in the doubled space through a path integral~\cite{Park:2017snt}, and  accordingly  string worldsheet actions which are fully covariant   with respect to symmetries like $\ODD$ T-duality, Weyl symmetry, target as well as worldsheet diffeomorphisms~\cite{Lee:2013hma,Arvanitakis:2017hwb,Arvanitakis:2018hfn}~(\textit{c.f.~}~\cite{Duff:1989tf,Tseytlin:1990nb,Tseytlin:1990va,
Hull:2004in,Hull:2006qs,Hull:2006va}), and  $\kappa$-symmetry~\cite{Park:2016sbw} (\ref{stringL}).\\
~\\
{$\bullet$ \textbf{Stringy graviton fields (closed-string massless NS-NS sector) as represented by $\left\{d,V_{Mp},\brV_{N\brq\,}\right\}$:}} \\
The defining properties of the DFT metric are
\be
\ba{ll}
{{\cal H}}_{MN}={{\cal H}}_{NM}\,,\qquad \quad
{{\cal H}}_{K}{}^{L} {{\cal H}}_{M}{}^{N} {{\cal J}}_{LN}={{\cal J}}_{KM}\,,
\ea
\label{defH}
\ee
from  which one can set  a pair of symmetric and orthogonal projectors,  
\[
\ba{ll}
P_{MN}=P_{NM}=\half(\cJ_{MN}+\cH_{MN})\,,\quad& \quad P_{L}{}^{M}P_{M}{}^{N}=P_{L}{}^{N}\,,\\
\brP_{MN}=\brP_{NM}=\half(\cJ_{MN}-\cH_{MN})\,,\quad &\quad \brP_{L}{}^{M}\brP_{M}{}^{N}=\brP_{L}{}^{N}\,,\qquad P_{L}{}^{M}\brP_{M}{}^{N}=0\,.
\ea
\]
Taking the ``square roots" of the  projectors, we acquire a pair of DFT vielbeins,  
\[
\ba{ll}
P_{MN}=V_{M}{}^{p}V_{N}{}^{q}\eta_{pq}\,,\quad&\quad
\brP_{MN}=\brV_{M}{}^{\brp}\brV_{N}{}^{\brq}\breta_{\brp\brq}\,,
\ea
\]
satisfying their own defining properties, 
\[
\ba{lll}
V_{Mp}V^{M}{}_{q}=\eta_{pq}\,,\qquad&\qquad
\brV_{M\brp}\brV^{M}{}_{\brq}=\breta_{\brp\brq}\,,\qquad&\qquad
V_{Mp}\brV^{M}{}_{\brq}=0\,,
\ea
\]
which are --- as the left inverse   of a matrix coincides with the right inverse --- equivalent to
\[
V_{M}{}^{p}V_{Np}+\brV_{M}{}^{\brp}\brV_{N\brp}=\cJ_{MN}\,.
\]
The most  general solutions to (\ref{defH}) can be  classified  by  two non-negative integers $(n,\bar{n})$~\cite{Morand:2017fnv},
\be
\cH_{MN}=\left(\ba{cc}H^{\mu\nu}&
-H^{\mu\sigma}B_{\sigma\lambda}+Y_{i}^{\mu}X^{i}_{\lambda}-
\brY_{\bri}^{\mu}\brX^{\bri}_{\lambda}\\
B_{\kappa\rho}H^{\rho\nu}+X^{i}_{\kappa}Y_{i}^{\nu}
-\brX^{\bri}_{\kappa}\brY_{\bri}^{\nu}\quad&~~
~~K_{\kappa\lambda}-B_{\kappa\rho}H^{\rho\sigma}B_{\sigma\lambda}
+2X^{i}_{(\kappa}B_{\lambda)\rho}Y_{i}^{\rho}
-2\brX^{\bri}_{(\kappa}B_{\lambda)\rho}\brY_{\bri}^{\rho}
\ea\right),
\label{cHFINAL}
\ee
where $1\leq i\leq n$,  $\,1\leq \bri\leq\brn$ and  
\[
\ba{lllll}
H^{\mu\nu}X^{i}_{\nu}=0\,,~~&~ H^{\mu\nu}\brX^{\bri}_{\nu}=0\,,~~&~K_{\mu\nu}Y^{\nu}_{i}=0\,,~~&~
K_{\mu\nu}\brY^{\nu}_{\bri}=0\,,~~&~H^{\mu\rho}K_{\rho\nu}+Y_{i}^{\mu}X^{i}_{\nu}
+\brY_{\bri}^{\mu}\brX^{\bri}_{\nu}=\delta^{\mu}_{~\nu}\,.
\ea
\]  
The corresponding  coset is, with $D=t+s+n+\brn$,  
\[
\frac{\ODD}{\mathbf{O}(t+n,s+n)\times\mathbf{O}(s+\brn,t+\brn)}\,,
\]
which has the dimension, $D^{2}-(n-\brn)^{2}$~\cite{priviateDC}, while $\cH_{M}{}^{M}=2(n-\brn)$ is $\ODD$ invariant.\\

\noindent Upon the generic $(n,\brn)$ background,  strings  become  chiral and anti-chiral over the $n$ and $\brn$ directions: 
\[
\ba{ll}
X^{i}_{\mu}\partial_{+}x^{\mu}=0\,,\qquad&\qquad \brX^{\bri}_{\mu}\partial_{-}x^{\mu}=0\,.
\ea
\]
Examples include  Riemannian  geometry as  $(0,0)$  where ${K_{\mu\nu}=g_{\mu\nu}}$, ${H^{\mu\nu}=g^{\mu\nu}}$,   Newton--Cartan gravity  as $(1,0)$,   Gomis--Ooguri  or Newton--Cartan non-relativistic strings as $(1,1)$~\cite{Gomis:2000bd,Harmark:2017rpg,Harmark:2018cdl,Berman:2019izh},     Carroll gravity  as $(D-1,0)$, and  Poisson--Lie dual $(1,1)$  backgrounds~\cite{Sakatani:2019jgu}.   In particular, the extreme case of $(D,0)$ corresponds to the maximally non-Riemannian, perfectly  $\ODD$ symmetric, vacuum geometry of  DFT, where the DFT metric coincides with the $\ODD$ metric, $\cH_{AB}=\cJ_{AB}$.  Intriguingly then, the  Riemannian as well as   partially non-Riemannian, ${n+\brn}<D$,   spacetimes   `emerge' after  spontaneously  breaking the $\ODD$ symmetry  with the  component fields in (\ref{cHFINAL}) interpreted  as    Goldstone bosons~\cite{Berman:2019izh}.  Furtheremore,  the maximally non-Riemannian  $(D,0)$ background does not allow any linear fluctuation: from the defining property~(\ref{defH}),   any linear fluctuation of the DFT metric must satisfy   $\delta\!\cH_{A}{}^{B}\cH_{B}{}^{C}+\cH_{A}{}^{B}\delta\!\cH_{B}{}^{C}=0$, and thus if $\cH_{AB}=\cJ_{AB}\,$, we have $\delta\cH_{AB}=0$\,.   Thus, taken as an internal space,  it realizes a graviscalar-moduli-free  Scherk--Schwarz twistable Kaluza--Klein reduction of DFT, in fact,  to  heterotic supergravity~\cite{Cho:2018alk}  .  \\
\\
{$\bullet$ \textbf{Covariant derivative:}}\\
The `master' covariant derivative,  
\[
\cD_{A}=\partial_{A}+\Gamma_{A}+\Phi_{A}+\brPhi_{A}\,,
\]
is characterized by  compatibilities  with the whole NS-NS sector,
\[
\ba{lll}
\cD_A{d} =0\,,\qquad&\qquad \cD_A V_{Bp} =0\,,\qquad&\qquad
\cD_{A} \bar{V}_{B\bar{p}} =0\,,
\ea
\]
as well as with the kinematical constant metrics,
\[
\ba{lllll}
\cD_{A}\cJ_{BC}=0\,,\quad&\qquad\cD_{A}\eta_{pq}=0\,,\quad&\qquad
\cD_{A}\breta_{\brp\brq}
=0\,,\quad&\qquad\cD_{A}C_{\alpha\beta}=0\,,\quad&\qquad\cD_{A}\brC_{\bralpha\brbeta}=0\,.
\ea
\]
The  DFT-Christoffel symbols are~\cite{Jeon:2011cn}
\[
\ba{ll}
\Gamma_{CAB}=&2\left(P\partial_{C}P\brP\right)_{[AB]}
+2\left({{\brP}_{[A}{}^{D}{\brP}_{B]}{}^{E}}-{P_{[A}{}^{D}P_{B]}{}^{E}}\right)\partial_{D}P_{EC}\\
{}&-4\left(\textstyle{\frac{1}{P_{M}{}^{M}-1}}P_{C[A}P_{B]}{}^{D}+\textstyle{\frac{1}{\brP_{M}{}^{M}-1}}\brP_{C[A}\brP_{B]}{}^{D}\right)\!\left(\partial_{D}d+(P\partial^{E}P\brP)_{[ED]}\right)\,,
\ea
\label{Gammao}
\]
and  the spin  connections are  
\[
\Phi_{Apq}=V^{B}{}_{p}(\partial_{A}V_{Bq}+{\Gamma_{AB}{}^{C}V_{Cq}})\,,\qquad\qquad\quad
\brPhi_{A\brp\brq}=\brV^{B}{}_{\brp}(\partial_{A}\brV_{B\brq}+\Gamma_{AB}{}^{C}\brV_{C\brq})\,.
\]
\\
\noindent {In Stringy Gravity there are no normal coordinates  where $\Gamma_{CAB}$ would vanish  point-wise: the Equivalence Principle  does not  hold  for strings, or extended objects.   However, when the  formalism is applied and restricted to  the case of  point particles, $\Gamma_{CAB}$ reduces to the ordinary  Christoffel symbols    and the Equivalence Principle is restored.  }   \\
~\\
{$\bullet$ \textbf{Scalar and `Ricci' curvatures:}}\\
\noindent The semi-covariant Riemann curvature in Stringy Gravity  is defined by 
\[
S_{ABCD}:=\half\left(R_{ABCD}+R_{CDAB}-\Gamma^{E}{}_{AB}\Gamma_{ECD}\right)\,,
\label{RiemannS}
\]
where $R_{C}{}_{DAB}=\partial_{A}\Gamma_{B}{}_{C}{}_{D}-\partial_{B}\Gamma_{A}{}_{C}{}_{D}+\Gamma_{A}{}_{C}{}_{E}\Gamma_{B}{}^{E}{}_{D}-\Gamma_{B}{}_{C}{}_{E}\Gamma_{A}{}^{E}{}_{D}$ (the ``field strength" of $\Gamma_{CAB}$).\\
The completely covariant `Ricci' and scalar curvatures are, with $S_{AB}=S_{ACB}{}^{C}$,  
\[
\ba{ll}
S_{p\brq}:=V^{A}{}_{p}\brV^{B}{}_{\brq}S_{AB}\,,\quad&\quad
\So:=\left(P^{AC}P^{BD}-\brP^{AC}\brP^{CD}\right)S_{ABCD}\,.
\ea
\label{RicciSc}
\]
~\\
{$\bullet$ \textbf{DFT  minimally coupled to matter:}}\\
While $e^{-2d}\So$ corresponds to the original DFT Lagrangian density~\cite{Siegel:1993xq,Hull:2009mi}, or the `pure' Stringy Gravity, the master covariant derivative fixes its minimal coupling  to extra matter fields, \textit{e.g.~}type II ${D=10}$  maximally supersymmetric  DFT~\cite{Jeon:2012hp},
\be
\ba{lll}
\cL_{{\scriptstyle{\rm{type\,II}}}}&=&e^{-2d}\left[\ba{l}
\textstyle{\frac{1}{8}}\So+\half\Tr(\cF\bar{\cF})
+i\brrho\cF\rhop
+i\brpsi_{\brp}\gamma_{q}\cF\brgamma^{\brp}\psip{}^{q}
+i\half\brrho \gamma^{p}\cD_{p}\rho
-i\half\brrhop \brgamma^{\brp}\cD_{\brp}\rhop\\
-i\brpsi^{\brp}\cD_{\brp}\rho
-i\half\brpsi^{\brp}\gamma^{q}\cD_{q}\psi_{\brp}
+i\brpsip{}^{p}\cD_{p}\rhop
+i\half\brpsip{}^{p}\brgamma^{\brq}\cD_{\brq}\psip_{p}
\ea\right]\,,
\ea
\label{L2}
\ee
or the Standard Model coupled to DFT~\cite{Choi:2015bga},
\be
\ba{lll}
\cL_{{\scriptstyle{\rm{SM}}}}&=&
e^{-2d}\left[\ba{l}
\frac{1}{16\pi G_{N}}\So+\sum_{\cV}\Tr(\cF_{p\brq}\cF^{p\brq})
+\sum_{\psi}\brpsi\gamma^{a}\cD_{a}\psi +\sum_{\psi^{\prime}}\brpsi^{\prime}\brgamma^{\bra}\cD_{\bra}\psi^{\prime}\\
-\cH^{AB}(\cD_{A}\phi)^{\dagger}\cD_{B}\phi\,-\,V(\phi)\,
+y_{d\,}\brq{\cdot\phi}\, d+y_{u\,}\brq{\cdot\tilde{\phi}}\, u+y_{e\,}\brl^{\prime}{\cdot\phi}\,e^{\prime}
\ea\right]\,.
\ea
\label{LSM}
\ee
The former Lagrangian~(\ref{L2})  was constructed to the full  \textit{i.e.~}quartic order  in fermions. It unifies  IIA and   IIB supergravities  as well as   ``Gomis--Ooguri supergravity"    as   different solution sectors.   The latter Lagrangian~(\ref{LSM})   may put quarks and leptons in  two  distinct spin group sectors, \textit{i.e.~}$\Spin(1,3)\,$ \textit{vs.} $\Spin(3,1)$. Every  single term in the above two Lagrangians is completely invariant with respect to the diffeomorphisms, twofold local Lorentz symmetries, and $\ODD$ T-duality.


\section{{Derivation of   the Einstein Double Field Equations}}
\noindent We consider  a general  action for Stringy Gravity (\textit{i.e.~}DFT) coupled to generic matter fields, $\Upsilon_{a}$, for example (\ref{L2}), (\ref{LSM}). The variation of the action  gives
\[
\ba{l}
\quad\,\delta\!\dis{\int e^{-2d}\Big[\,\textstyle{\frac{1}{16\pi G}}\So+L_{\rm{matter}}\,\Big]}\\
= \dis{\int e^{-2d}\left[\textstyle{\frac{1}{4\pi G}}
\brV^{A\brq}\delta V_{A}{}^{p}(S_{p\brq}-8\pi G\EK_{p\brq})-\textstyle{\frac{1}{8\pi G}}\delta d(\So-8\pi G\To)+
\delta\Upsilon_{a}\dis{\frac{\delta L_{\rm{matter}}}{\delta \Upsilon_{a}}}\right]}\\
=\dis{\int e^{-2d}\left[
\textstyle{\frac{1}{8\pi G}}\xi^{B}\cD^{A}\left\{ G_{AB} - 8 \pi G T_{AB} \right\}+
(\hcL_{\xi}\Upsilon_{a})\,\dis{\frac{\delta L_{\rm{matter}}}{\delta \Upsilon_{a}}}\right]}\,, \label{EQ13}
\ea
\]
where the second line is for generic variation and the third line is specifically for diffeomorphic transformation.  While deriving the above,  one is  naturally led to define
\[
\ba{ll}
\EK_{p\brq}:=\dis{\frac{1}{2} \left(V_{Ap}\frac{\delta L_{\rm{matter}}}{\delta \brV_{A}{}^{\brq}}-\brV_{A\brq}\frac{\delta L_{\rm{matter}}}{\delta V_{A}{}^{p}}\right)\,,}\qquad&\qquad
\To:= e^{2d}\times\dis{\frac{\delta\left(e^{-2d} L_{\rm{matter}}\right)}{\delta d}\,,}
\ea
\label{cK}
\]
and subsequently  also the stringy  \textit{Einstein curvature}, $G_{AB}$, and \textit{Energy-Momentum tensor}, $T_{AB\,}$, 
\[
\ba{ll}
G_{AB}=4V_{[A}{}^{p}\brV_{B]}{}^{\brq}S_{p\brq}-\half\cJ_{AB}\So\,,\qquad&\quad
\cD_{A}G^{AB}=0\qquad(\mbox{off-shell})\,, \\ 
\EM_{AB}:=4V_{[A}{}^{p}\brV_{B]}{}^{\brq}\EK_{p\brq}-\half\cJ_{AB}\To\,,\qquad&\quad
\cD_{A}\EM^{AB}= 0\qquad(\mbox{on-shell})\,,
\ea
\]
which satisfy  $G_{A}{}^{A}=-D\So$, $T_{A}{}^{A}=-D\To$. Therefore, 
the equations of motion of the stringy graviton fields  are  unified into a single expression,    \textit{the Einstein Double Field Equations}~(\ref{EDFE}). \\
%

\noindent Restricting to the $(0,0)$ Riemannian background, the Einstein Double Field Equations  reduce to 
\begin{eqnarray}
R_{\mu\nu}+2\trd_{\mu}(\partial_{\nu}\phi)-\quarter H_{\mu\rho\sigma}H_{\nu}{}^{\rho\sigma}
&=&8\pi G\EK_{(\mu\nu)}\,,\nonumber\\
\trd^{\rho}\!\left(e^{-2\phi}H_{\rho\mu\nu}\right)&=&16\pi Ge^{-2\phi}\EK_{[\mu\nu]}\,,\nonumber\\
R+4\Box\phi-4\partial_{\mu}\phi\partial^{\mu}\phi-\textstyle{\frac{1}{12}}H_{\lambda\mu\nu}H^{\lambda\mu\nu}&=&8\pi G\To\,,\nonumber
\end{eqnarray}
which imply the conservation law,  $\cD_{A}\EM^{AB} = 0$,  now given explicitly by
\[
\ba{ll}
\na^{\mu}\EK_{(\mu\nu)}-2\partial^{\mu}\phi\,\EK_{(\mu\nu)}+\half H_{\nu}{}^{\lambda\mu}\EK_{[\lambda\mu]}-\half\partial_{\nu}\To 
=0\,,\quad&\quad
\na^{\mu}\!\left(e^{-2\phi}\EK_{[\mu\nu]}\right)=0\,.
\ea
\]
The Einstein Double Field Equations also govern the dynamics of other non-Riemannian cases,  $(n,\brn)\neq(0,0)$, where  the invertible  Riemannian metric, $g_{\mu\nu}$, cannot be defined.

\section*{{Examples}}
\vspace{-5pt}
\begin{itemize}
\item[--]  \textit{Pure Stringy Gravity with the $\ODD$ invariant  cosmological constant:}
\[
\ba{lll}
\textstyle{\frac{1}{16\pi G}}e^{-2d}\left(\So-2\Lambda_{\DFT}\right)\,,\quad&\quad
\EK_{p\brq}=0\,, \quad &\quad\To=\frac{1}{4\pi G}\Lambda_{\DFT}\,.
\ea
\]
 
\item[--] \textit{RR sector,  represented  by  a $\Spint\times\oSpint$ bi-spinorial potential, $\cC^{\alpha}{}_{\bralpha}$\,:}
\[
\ba{lll}
L_{\mathrm{RR}}=\half \Tr\left(\cF\brcF\right)\,,\quad&\quad
\EK_{p\brq}=-\quarter\Tr\left(\gamma_{p}\cF\brgamma_{\brq}\brcF\right)\,,\quad&\quad\To=0\,,
\ea
\]
where   
\[
\cF=\cD_{+}\cC=
\gamma^{p}\cD_{p}\cC+\gammae\cD_{\brp}
\cC\brgamma^{\brp}\,,
\]
which is the RR flux set by  an $\ODD$ covariant   `$H$-twisted' cohomology, $(\cD_{+})^{2}=0$, and $\brcF=\brC^{-1}\cF^{T}C$ is its charge conjugate~\cite{Jeon:2012hp}.  

\item[--] \textit{Scalar field\,:} 
\[
\ba{lll}
L_{\Phi}=-\half\cH^{MN}\partial_{M}\Phi\partial_{N}\Phi-\half\mPhi^{2}\Phi^{2}\,,\quad&\quad
\EK_{p\brq}=\partial_{p}\Phi\partial_{\brq}\Phi\,,\quad&\quad\To=-2L_{\Phi}\,.
\ea
\]

\item[--] \textit{Spinor field:} 
\[
\ba{lll}
L_{\psi}=\brpsi\gamma^{p}\cD_{p}\psi+\mpsi\brpsi\psi\,,\quad&\quad
\EK_{p\brq}=-\quarter(\brpsi\gamma_{p}\cD_{\brq}\psi-\cD_{\brq}\brpsi\gamma_{p}\psi)\,,
\quad&\quad\To= 0\,.
\ea
\]

\item[--] \textit{Point particle:\,}
\be
e^{-2d}L_{\rm{particle}}=\int\rd\tau~\big[\,e^{-1\,}D_{\tau}y^{A}D_{\tau}y^{B}\cH_{AB}(x)-\quarter m^{2}e\,\big]\delta^{D\!}\big(x-y(\tau)\big)\,,
\label{particleL}
\ee
\vspace{-15pt}
\[
\ba{ll}
\EK_{p\brq}=-\dis{\int\rd\tau}~2e^{-1\,}(D_{\tau}y^{A}V_{Ap})(D_{\tau}y^{B}\brV_{B\brq})\,e^{2d(x)}\delta^{D\!}\big(x-y(\tau)\big)\,,
\quad&\quad\To=0\,.
\ea
\]

\item[--] \textit{Green-Schwarz superstring ($\kappa$-symmetric)\,:}
\be
e^{-2d}L_{\rm{string}}={\textstyle{\frac{1}{4\pi\alpha^{\prime}}}}{\dis{\int}}\rd^{2}\sigma\left[-\half\sqrt{-h}h^{ij}\Pi_{i}^{M}\Pi_{j}^{N}\cH_{MN}-\epsilon^{ij}D_{i}y^{M}(\cA_{j M}-i\Sigma_{jM})\right]\delta^{D\!}\big(x-y(\sigma)\big)\,,
\label{stringL}
\ee
\vspace{-15pt}
\[
\ba{ll}
\EK_{p\brq}(x)={\textstyle{\frac{1}{4\pi\alpha^{\prime}}}}{\dis{\int}}\rd^{2}\sigma\sqrt{-h}h^{ij}(\Pi_{i}^{M}V_{Mp})(\Pi_{j}^{N}\brV_{N\brq})\,e^{2d}\delta^{D\!}\big(x-y(\sigma)\big)\,,
\quad&\quad\To=0\,,
\ea
\]
where\,  $\Sigma_{i}^{M}=\brtheta\gamma^{M}\partial_{i}\theta+
\brtheta^{\prime}\brgamma^{M}\partial_{i}\theta^{\prime}\,$ and\, $\Pi_{i}^{M}=\partial_{i}y^{M}-\cA_{i}^{M}-i\Sigma_{i}^{M}$~\cite{Park:2016sbw}.
\end{itemize}

\section{DFT as Modified Gravity}
\noindent As DFT evolves into Stringy Gravity,  which appears  to, at least conceptually,  differ  from General Relativity, it  should be  of interest, if not the duty of  a  physicist,  to  investigate  how DFT  as a graviational theory modifies GR.  Since the stringy Energy-Momentum tensor  has $D^{2}+1$   components, and this   is certainly  larger  than $\half D(D+1)$ which is the number of  components  in GR,   it is natural to expect that  the gravitational phenomena are richer in Stringy Gravity than in General Relativity.    As a first step to verify  this,  henceforth we  focus on   the most general,  static,  spherically symmeric, asymtotically flat,  Riemannian,  regular `star-like' solution to the ${D=4}$ Einstein Double Field Equations,
\be
G_{AB}=\left\{\,\ba{cllc}8\pi G T_{AB}&\quad\mbox{for}\quad&r\leq\rc\quad&\mbox{(inside~the~stringy~star)}\\
0&\quad\mbox{for}\quad&r> \rc\quad&\mbox{(ourside)\,.}\ea\right.
\label{EDFEsph}
\ee
Outside the stringy star, we have the spherical  `vacuum'  geometry~\cite{Burgess:1994kq,Ko:2016dxa},
\be
{\ba{lc}
e^{2\phi}=\gamma_{+}\left(\frac{r-\alpha}{r+\beta}\right)^{\frac{b}{\sqrt{a^{2}+b^{2}}}}+\gamma_{-}\left(\frac{r+\beta}{r-\alpha}\right)^{\frac{b}{\sqrt{a^{2}
+b^{2}}}}\,,\qquad&\qquad
H_{\scriptscriptstyle{(3)}}=h\sin\vartheta\,\rd t\wedge\rd\vartheta\wedge\rd\varphi\,,\\
\multicolumn{2}{c}{
\rd s^{2}=e^{2\phi}\left[-\left(\frac{r-\alpha}{r+\beta}\right)^{\frac{a}{\sqrt{a^{2}+b^{2}}}}\rd t^{2}
+\left(\frac{r+\beta}{r-\alpha}\right)^{\frac{a}{\sqrt{a^{2}+b^{2}}}}\big\{\rd r^{2}+(r-\alpha)(r+\beta)
\rd\Omega^{2}\big\}\right]\,,}
\ea}
\ee
where $a,b,h,\alpha,\beta$ are  constant parameters satisfying the constraint,   $a^{2}+b^{2}=(\alpha+\beta)^{2}\,$; we let $\gamma_{\pm}:=\half(1\pm\sqrt{1-h^{2}/b^{2}})\,$; and  $\rmd s^{2}$ is given  in string frame. Thus there are    four independent  free parameters in the spherical vacuum geometry, in contrast to the Schwarzschild geometry which possesses   only  one free parameter, \textit{i.e.~}mass.  The Einstein Double Field Equations~(\ref{EDFEsph}) then  determine --- and hence reveal the physical meaning of --- all  of these   ``free" parameters in terms of    the   Energy-Momentum tensor  inside the stringy star,   for example, 
\[
a={\dis{{\int_{0}^{\red{\rc}}\!\rd r}\int_{0}^{\pi}\!\rd\vartheta\int_{0}^{2\pi}\!\rd\varphi}}~e^{-2d}\Big[\textstyle{\frac{1}{4\pi}}H_{r\vartheta\varphi}H^{r\vartheta\varphi}+2 G\left(\EK_{r}{}^{r}+\EK_{\vartheta}{}^{\vartheta}+\EK_{\varphi}{}^{\varphi}-\EK_{t}{}^{t}-\To\right)\Big]\,.
\]
The $\ODD$ symmetric doubled-yet-gauged  particle action~(\ref{particleL})  implies that a point-like particle should follow  a geodesic defined in  string frame~\cite{Ko:2016dxa}, rather than in Einstein frame.\footnote{However, this is not an S-duality invariant statement. The author would like to thank Chris Hull for this remark. Our discussion is thus restricted  to the implications of $\ODD$ T-duality rather than S- or U-dualities. }
In terms of   the areal radius, $R$, which normalizes the angular part of the metric,  $\rmd s^{2}=g_{tt}\rmd t^{2}+g_{RR}\rmd R^{2}+R^{2}\rmd\Omega^{2}$,  the orbital velocity of a point particle probe  can be computed from
\[
V_{\rm{orbit}}=\sqrt{R\,\frac{\rmd  \Phi_{\rm{Newton}}}{\rmd R\qquad}}\,,
\]
\[
\Phi_{{\rm{Newton}}}=-\half(1+g_{tt})= -\frac{MG}{R}+\left(\frac{\,2b^{2}-h^{2}+2ab\sqrt{1-h^{2}/b^{2}}\,}{\,a^{2}+b^{2}-h^{2}+2ab\sqrt{1-h^{2}/b^{2}}\,}\right)
\left(\frac{MG}{R}\right)^{2}\,+\,\cdots~ \,,
\label{PhiNewton}
\]
where  the ellipses in (\ref{PhiNewton}) denote   higher order terms in $\frac{MG}{R}$, and the mass is given by
\be
\textstyle{MG=\frac{1}{2}\!\left(a+ b\sqrt{1-h^{2}/b^{2}}\right)=}
{\dis{{\int_{0}^{\red{\infty}}\!\rd r}\int_{0}^{\pi}\!\rd\vartheta\int_{0}^{2\pi}\!\rd\varphi}~e^{-2d}\left(-2G\EK_{t}{}^{t}+\textstyle{\frac{1}{8\pi}}\left|H_{t\vartheta\varphi}H^{t\vartheta\varphi}\right|
 \right)}\,.
\label{Minfty}
\ee
Thus, in terms of the dimensionless radius,  $R/(MG)$, normalized by the mass times the Newton constant, the orbital motion  becomes Keplerian, \textit{i.e.~}$V_{\rm{orbit}}\simeq \sqrt{\frac{MG}{R}}$,  for large  $R/(MG)$, while it is  non-Keplerian for small $R/(MG)$.  That is to say, Stringy Gravity modifies General Relativity at ``short"  dimensionless scales.  In fact, depending on the parameters, the gravitational force can even be repulsive at ``short" scale.  This might shed new light upon  the dark matter/energy problems, as they   arise  essentially from  ``short" dimensionless scale  observations:
\vspace{-2pt}
\begin{center}
\includegraphics[width=15.1cm]{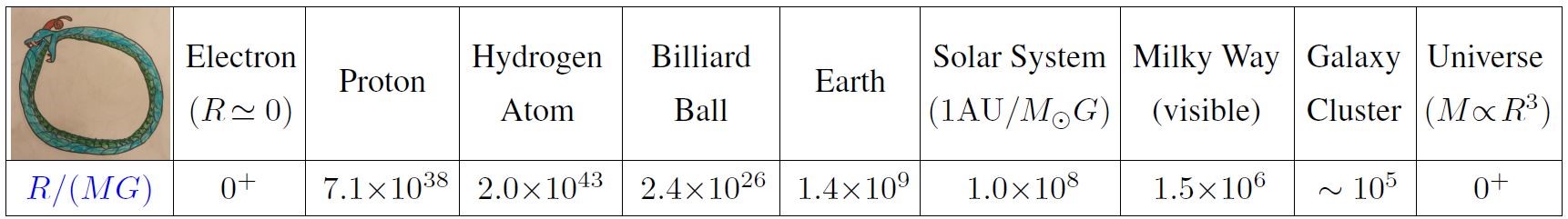}
\small{\textbf{`Uroboros' spectrum of the dimensionless radial variable normaized by mass in natural units~\cite{Ko:2016dxa,Park:2017snt}. \\The  observations of stars and galaxies  far away may reveal   the short-distance nature of gravity.\\
Repulsive gravitational  force  at short scale may explain the acceleration of the Universe.   }}
\vspace{2pt}
\end{center}

\noindent Finally, we speculate  that   electric  $H$-flux  may be   \textit{dark matter}, since it contributes to the mass formula~(\ref{Minfty})  while it    decouples   from   point particles~(\ref{particleL}).  We call for verification. \\

\section*{Dedication}
\noindent I would like to dedicate this humble  writing  to the memory of Cornelius Sochichiu who has taught me how to  balance life and physics until his last moment.
\section*{Acknowledgements}
\noindent I would like  to thank  the organizers of   Corfu Summer Institute 2018 as well as subsequent   meetings,  \textit{Double Field Theory: Progress and Applications}   at  University of Cape Town,  \textit{String: T-duality, Integrability and Geometry} at  Tohoku University, and \textit{100+4 General Relativity and Beyond} at Jeju National University  supported by  APCTP. Therein  I have benefitted   from  stimulating  discussions, among others, with  David Berman, Chris Blair, Robert  Brandenberger,  Chris Hull, Ctirad Klim{\v c}{\' i}k,  Kanghoon Lee, Yuho Sakatani,  and Satoshi Watamura.  This work  was  supported by  the National Research Foundation of Korea   through  the Grant  NRF-2016R1D1A1B01015196.\\

\end{document}